\documentclass[aps,pra,showpacs, twocolumn]{revtex4}
\usepackage{graphicx}    
\usepackage{amsmath}

\begin{document}

\title{Field-quadrature and photon-number variances for Gaussian states}

\author{Moorad Alexanian}
\email[]{alexanian@uncw.edu}

\affiliation{Department of Physics and Physical Oceanography\\
University of North Carolina Wilmington\\ Wilmington, NC
28403-5606\\}

\date{\today}

\begin{abstract}
We calculate exactly the quantum mechanical, temporal characteristic function $\chi(\eta)$ for a single-mode, degenerate parametric amplifier for a system in the Gaussian state, viz., a displaced-squeezed thermal state. Knowledge of $\chi(\eta)$ allows only the determination of the time development of arbitrary functions of equal-time products of creation and annihilation photon operators. We calculate, in particular, the fluctuations in photon number, quadrature operators, and quadrature variance. We contrast the very important difference between the nonclassicality criteria based on the one-time characteristic function $\chi(\eta)$ versus nonclassicality criteria based on the two-time, second-order coherence function $g^{(2)}(\tau)$ and show numerically that the nonclassicality criteria based on $\chi(\eta)$ does not determine the classical/nonclassical behavior of $g^{(2)}(\tau)$.
\end{abstract}

\maketitle {}
\section{Introduction}

The generation of nonclassical radiation fields, e.g., quadrature-squeezed light, photon antibunching, sub-Poissonian statistics, etc., establishes the discrete nature of light and serves to study fundamental questions regarding the interaction of quantized radiation fields with matter \cite{GSA13}.

In a recent work \cite{MA16}, a detailed study was made of the temporal development of the second-order coherence function for Gaussian states--displaced-squeezed thermal states---the dynamics governed by a Hamiltonian for degenerate parametric amplification. The time development of the Gaussian state is generated by an initial thermal state and the system subsequently evolves in time where the usual assumption of statistically stationary fields is not made. Nonclassicality were observed for various values of the parameters governing the temporal development of the coherence function $g^{(2)}(\tau)$---such as the coherent parameter $\alpha$, squeeze parameter $\xi$, and the mean photon number $\bar{n}$ of the initial thermal state. Our characterization of nonclassicality was based solely on the coherence function violating inequalities satisfied by the classical correlation functions. 

In the present work we dwell into the notion of nonclassicality based on using the characteristic function to study fluctuations in photon number, quadrature operators, and quadrature variance. In Section 2, we consider the general Hamiltonian of the degenerate parametric amplifier (DPA). Section 3 deals with the characteristic function and the field-quadrature variance. In Section 4, we obtain the photon-number variance. Section 5 deals with differing characterization of nonclassicality. In Section 6, we study numerical examples to elucidate the temporal behavior of differing quantities in order to study the question of necessary and sufficient conditions for nonclassicality. Finally, Section 7 summarizes our results.

\section {Degenerate parametric amplification}

The Hamiltonian for degenerate parametric amplification, in the interaction picture, is
\begin{equation}
\hat{H} = c \hat{a}^{\dag 2} + c^* \hat{a}^2 + b\hat{a} + b^* \hat{a}^\dag.
\end{equation}
The system is initially in a thermal state $\hat{\rho}_{0}$ and a after a preparation time $t$, the system temporally develops into a Gaussian state and so \cite{MA16}
\begin{equation}
\hat{\rho}_{G}=\exp{(-i\hat{H}t/\hbar)}\hat{\rho}_{0} \exp{(i\hat{H}t/\hbar)}
\end{equation}
\[
= \hat{D}(\alpha) \hat{S}(\xi)\hat{\rho}_{0} \hat{S}(-\xi) \hat{D}(-\alpha),
\]
with  the displacement $\hat{D}(\alpha)= \exp{(\alpha \hat{a}^{\dag} -\alpha^* \hat{a})}$  and the squeezing $\hat{S}(\xi)=  \exp\big{(}-\frac{\xi}{2} \hat{a}^{\dag 2} + \frac{\xi^*}{2} \hat{a}^{2} \big{ )}$ operators, where $\hat{a}$ ($\hat{a}^{\dag})$ is the photon annihilation (creation) operator, $\xi = r \exp{(i\theta)}$, and $\alpha= |\alpha|\exp{(i\varphi)}$. The thermal state is given by

\begin{equation}
\hat{\rho}_{0} = \exp{(-\beta \hbar \omega\hat{n})}/ \textup{Tr}[\exp{(-\beta \hbar \omega \hat{n})}],
\end{equation}
with $\hat{n}= \hat{a}^\dag \hat{a}$ and $\bar{n}= \textup{Tr}[\hat{\rho}_{0} \hat{n}]$ .

The parameters $c$ and $b$ in the degenerate parametric Hamiltonian (1) are determined \cite{MA16} by the parameters  $\alpha$ and $\xi$ of the Gaussian density of state (2) via
\begin{equation}
tc = -i\frac{\hbar}{2} r\exp(i\theta)
\end{equation}
and
\begin{equation}
tb= -i\frac{\hbar}{2}\Big{(} \alpha \exp{(-i\theta)} + \alpha^* \coth (r/2)\Big{)} r,
\end{equation}
where $t$ is the time that it takes for the system governed by the Hamiltonian (1) to generate the Gaussian density of state $\hat{\rho}_{G}$ from the initial thermal density of state $\hat{\rho}_{0}$.

The quantum mechanical seconde-order, degree of coherence is given by
\begin{equation}
g^{(2)}(\tau) = \frac{\langle \hat{a}^{\dag}(0) \hat{a}^{\dag}(\tau) \hat{a}(\tau) \hat{a}(0)\rangle }{\langle \hat{a}^{\dag}(0)  \hat{a}(0)\rangle \langle \hat{a}^{\dag}(\tau)\hat{a}(\tau) \rangle},
\end{equation}
where all the expectation values are traces with the Gaussian density operator, viz., a displaced-squeezed thermal state. Accordingly, the system is initially in the thermal state $\hat{\rho}_{0}$. After time $t$, the system evolves to the Gaussian state $\hat{\rho}_{G}$ and a photon is annihilated at time $t$, the system then develops in time and after a time $\tau$ another photon is annihilated \cite{MA16}. Therefore, two photon are annihilated in a time separation $\tau$ when the system is in the Gaussian density state $\hat{\rho}_{G}$.

It is important to remark that we do not suppose statistically stationary fields. Therefore, owing to the $\tau$ dependence of the number of photons in the cavity in the denominator of Equation (6), the system asymptotically, as $\tau\rightarrow \infty$, approaches a finite limit without supposing any sort of dissipative processes \cite{MA16}. The coherence function $g^{(2)}(\tau)$ is a function of $\Omega \tau=(r/t)\tau$, $\alpha$, $\xi$, and the average number of photons $\bar{n}$ in the initial thermal state (3), where the preparation time $t$ is the time that it takes the system to dynamically generate the Gaussian density $\hat{\rho}_{G}$ given by (2) from the initial thermal state $\hat{\rho}_{0}$ given by (3). Note that the limit $r\rightarrow 0$ is a combined limit whereby $\Omega =r/t$ also approaches zero resulting in a correlation function which has a power law decay as $\tau/t \rightarrow \infty$ rather than an exponential law decay as $\tau/t \rightarrow \infty$ as is the case in the presence of squeezing when $r>0$ \cite{MA16}.

\section{characteristic function: Field-quadrature variance}

The calculation of the correlation function (6) deals with the measurement of observables at two different times. On the other hand, a complete statistical description of a field involves the expectation value of any function of the equal-time operators $\hat{a}$ and $\hat{a}^\dag$. A characteristic function contains all the necessary information to reconstruct the density matrix for the state of the field at one time but does no suffice to calculate the correlation $g^{(2)}(\tau)$, which involves two times rather than only one time as is the case with the characteristic function.

Now \cite{MA16}
\begin{equation}
\hat{\rho}(t+\tau) =\exp\big{(}-i\hat{H}(t+\tau)\big{)} \hat{\rho}_{0}\exp \big{(}i\hat{H}(t+\tau)\big{)}
\end{equation}
\[
= \exp(-i\hat{H}\tau) \hat{\rho}_{G}\exp(i\hat{H}\tau).
\]
Accordingly, for any operator function $\mathcal{\hat{O}}(\hat{a},\hat{a}^\dag)$, one has that
\begin{equation}
\textup{Tr}[\hat{\rho}(t+\tau) \mathcal{\hat{O}}(\hat{a},\hat{a}^\dag)] = \textup{Tr}[\hat{\rho}_{G} \mathcal{\hat{O}}\big{(}\hat{a}(\tau),\hat{a}^\dag (\tau)\big{)}]
\end{equation}
\[
\equiv \langle  \mathcal{\hat{O}}\big{(}\hat{a}(\tau),\hat{a}^\dag (\tau)\big{)} \rangle .
\]

One obtains for the characteristic function
\[
\chi(\eta) = \textup{Tr}[\hat{\rho}(t+\tau)\exp{(\eta \hat{a}^\dag}-\eta^*\hat{a})]\exp{(|\eta|^2/2)}
\]
\[
=\textup{Tr}[\hat{\rho}(t+\tau)\exp{(\eta \hat{a}^\dag}) \exp{(-\eta^*\hat{a})}]
\]
\begin{equation}
=\exp{(|\eta|^2/2)} \exp{\big{(}\eta A^*(\tau)- \eta^* A(\tau)\big{)}}\cdot
\end{equation}
\[
\cdot \exp{\big{(}-(\bar{n}+1/2)|\xi(\tau)|^2\big{)}},
\]
where

\[
A(\tau) =\alpha \Bigg{(}\cosh(\Omega\tau)+\frac{1}{2}\coth(r/2) \sinh (\Omega \tau)
\]
\begin{equation}
-\frac{1}{2} (\cosh(\Omega \tau)-1)+\exp[i(\theta -2 \varphi)]\Big{[} -\frac{1}{2}\sinh(\Omega\tau)
\end{equation}
\[
-\frac{1}{2}\coth(r/2)\big{(}\cosh(\Omega\tau)-1\big{)}\Big{]} \Bigg{)}
\]
and
\begin{equation}
\xi(\tau)= \eta\cosh(\Omega \tau +r) +\eta^* \exp(i\theta) \sinh(\Omega \tau +r).
\end{equation}
Define
\begin{equation}
|\xi(\tau)|^2 = \eta^2 T^*(\tau) +\eta^{*2} T(\tau) +\eta \eta^* S(\tau),
\end{equation}
with
\begin{equation}
T(\tau)= \frac{1}{2} \exp{(i \theta)} \sinh [2(\Omega \tau + r)]
\end{equation}
and
\begin{equation}
S(\tau)= \cosh[2(\Omega \tau +r)].
\end{equation}

With the aid of successive derivatives of the characteristic function $\chi(\eta)$, one obtains for the quadrature $\hat{x}_{\lambda}$ and the quadrature variance $\Delta x^2_{\lambda}$

\begin{equation}
\langle \hat{x}_{\lambda}\rangle = \textup{Tr}[\hat{\rho}(t+\tau) \hat{x}_{\lambda}]  =  \frac{1}{\sqrt{2}}[A(\tau) e^{-i\lambda} + A^*(\tau) e^{i\lambda}]
\end{equation}
and
\[
\Delta x^2_{\lambda} =\textup{Tr}[\hat{\rho}(t+\tau)( \hat{x}_{\lambda} -\langle \hat{x}_{\lambda}\rangle)^2 ]
\]
\begin{equation}
=(\bar{n} + 1/2) \Big{(}\exp{[2(\Omega \tau +r)]}\sin^2(\lambda-\theta/2)
\end{equation}
\[
+  \exp[-2(\Omega \tau +r)] \cos^2(\lambda-\theta/2) \Big{ )},
\]
where $\hat{x}_{\lambda} = (\hat{a}e^{-i \lambda} + \hat{a}^\dag e^{i \lambda})/\sqrt{2}$. The phase-sensitive quadrature operators represent a set of observables that can be measured for radiation modes, atomic motion in a trap, and other related systems \cite{WVO99}.

The expectation value of $\hat{x}_{\lambda}$ is determined by the coherent amplitude  $\alpha$ as well as the squeezing parameter $\xi$ while the variance $\Delta x^2_{\lambda}$, and hence the squeezing, depends on the squeezing parameter $\xi$ only. The product of the variances of the two quadratures components $\hat{x}_{\lambda}$ and $\hat{x}_{\lambda + \pi/2}$ is bounded from below by the Heisenberg uncertainty principle
\begin{equation}
\Delta x^2_{\lambda} \Delta x^2_{\lambda\ \pi/2}= (\bar{n}+1/2)^2  \big{(}\cosh^2[2(\Omega \tau +r)]
\end{equation}
\[
- \cos^2(\theta- 2 \lambda) \sinh^2[2(\Omega \tau +r)]\big{)}\geq (\bar{n}+1/2)^2 \geq  \frac{1}{4}.
\]

The signal-to-noise ratio \cite{RL00} is defined as
\begin{equation}
\textup{SNR} = \frac{(\langle \hat{x}_{\lambda}\rangle)^2}{\Delta x^2_{\lambda}}.
\end{equation}
Thus the maximum signal-to-noise ratio is
\begin{equation}
\textup{SNR}_{\textup{max}} =|\alpha|^2 \frac{\big{[} \coth(r/2)(1- e^{-\Omega \tau}) +(1 +e^{-\Omega \tau})\big{]}^2}{(2\bar{n} +1) e^{-2(\Omega \tau +r)}},
\end{equation}
for $\varphi =\lambda= \theta/2$. The result for the squeezed coherent state, $4 e^{2r}|\alpha|^2$, follows for $\tau=0$ and $\bar{n}=0$.

\section{ Photon-number variance}
The time development of the photon number is given by
\[
\textup{Tr}[\hat{\rho}(t+\tau)\hat{a}^\dag \hat{a}] = \langle \hat{a}^\dag(\tau) \hat{a}(\tau)\rangle = \langle\hat{n}(\tau)\rangle
\]
\begin{equation}
=(\bar{n}+1/2)\cosh[2(\Omega\tau+r)] + |A(\tau)|^2 -\frac{1}{2},
\end{equation}
while the variance is
\[
\Delta n^2(\tau) = \textup{Tr}[\hat{\rho}(t+\tau) (\hat{n} - \langle \hat{n}\rangle)^2] = (\bar{n}+1/2)^2 \cosh[4(\Omega \tau +r)]
\]
\begin{equation}
 +(\bar{n}+1/2)\Big{(} 2\cosh[2(\Omega \tau +r)]|A(\tau)|^2 -\sinh[2(\Omega \tau+r)]\cdot
\end{equation}
\[
\cdot[e^{i\theta} A^{*2}(\tau)+e^{-i\theta} A^2(\tau)]\Big{)} -\frac{1}{4}.
\]

Note, contrary to the quadrature variance (16), the photon-number variance (21) depends, in addition to the squeezing parameter $\xi$,  also on the coherent amplitude $\alpha$ via $A(\tau)$ given by Equation (10).

\section{Nonclassicality criteria}

Nonclassical light can be characterized differently, for instance, with the aid of the quantum degree of second-order coherence $g^{(2)}(\tau)$ by the nonclassical inequalities

\begin{equation}
g^{(2)}(0)< 1 \hspace{0.3in} \textup{and}    \hspace{0.3in} g^{(2)}(0) <  g^{(2)}(\tau),
\end{equation}
where the first inequality represents the sub-Poissonian statistics, or photon-number squeezing, while the second gives rise to antibunched light. Hence a measurement of $g^{(2)}(\tau)$ can be used to determine the nonclassicality of the field. The two nonclassical effects often occur together but each can occur in the absence of the other. Similarly, one can derive the nonclassical inequality \cite{RC88}
\begin{equation}
|g^{(2)}(0)-1| < |g^{(2)}(\tau)-1|,
\end{equation}
that is, $g^{(2)}(\tau)$ can be farther away from unity than it was initially at $\tau=0$.

In the Glauber-Sudarshan coherent state or P-representation of the density operator $\hat{\rho}$ one has that \cite{GSA13}
\begin{equation}
\hat{\rho} =\int P(\alpha)|\alpha\rangle \langle\alpha| \textup{d}^2\alpha,
\end{equation}
where $|\alpha\rangle$ is a coherent state and nonclassicality occurs when $P(\alpha)$ takes on negative values and becomes more singular than a Dirac delta function. One has the normalization condition $\int P(\alpha) \textup{d}^2\alpha=1$; however, $P(\alpha)$ would not describe probabilities, even if positive, of mutually exclusive states since coherent states are not orthogonal. In fact, coherent states are over complete.

A sufficient conditions for the nonclassicality is for the quadrature of the field to be narrower than that for a coherent state, that is,
\begin{equation}
\Delta x^2_{\lambda} < \frac{1}{2}.
\end{equation}

Another sufficient condition is determined by the Mandel $Q_{M}(\tau)$ parameter related to the photon-number variance \cite{GSA13}
\begin{equation}
Q_{M}(\tau)= \frac{\Delta n^2(\tau) -\langle \hat{n}(\tau)\rangle}{\langle \hat{n}(\tau)\rangle},
\end{equation}
where $-1 \leq Q_{M}(\tau) <0$ implies that $P(\alpha)$ assumes negative values and thus the field must be nonclassical with sub-Poissonian statistics. Condition $Q_{M}(0) <0$ is equivalent to the first condition in Equation (22) since $Q_{M}(0) = \langle \bar{n}(0)\rangle[g^{(2)}(0)-1]$. If, however, both the Mandel $Q_{M}$ parameter and the squeezing parameter $(\Delta x_{\lambda}^2-1/2)$ are positive, then no conclusion can be drawn on the nonclassical nature of the radiation field.

A purported necessary and sufficient condition for nonclassicality is
\begin{equation}
|\chi(\eta)|>1,
\end{equation}
which is actually only the lowest order of a hierarchy of conditions that must be satisfied for a quantum state to be nonclassical \cite{RV02}. One obtains for the characteristic function (9) of a Gaussian state that
\begin{equation}
|\chi(\eta)| =  \exp \Big{[}\frac{1}{2} |\eta|^2\Big{(}1-(2\bar{n}+1) e^{-2(\Omega \tau+r)}\Big{)}\Big{]},
\end{equation}
for $ 2\omega = \theta + \pi$, where $\eta= |\eta| e^{i \omega}$.
The nonclassicality condition (27) becomes
\begin{equation}
(2\bar{n}+1) e^{-2(\Omega \tau+r)}<1,
\end{equation}
which is the same as that given by condition (25) when $\theta= 2 \lambda$. If the nonclassicality condition (29) holds for $\tau=0$, then it holds for $\tau>0$. Accordingly, our dynamical system if initially nonclassical remains so as time goes on.

It is important to remark that all the criteria based on the characteristic function $\chi(\eta)$ are associated with knowledge of the dynamical system at a single time whereas criteria based on the degree of second-order coherence $g^{(2)}(\tau)$ are associate with measurements of an observable at two different times. Accordingly, the purported necessary and sufficient condition for nonclassicality given by (29), which does not depend on the coherent amplitude $\alpha$, may give rise to coherence functions $g^{(2)}(\tau)$, which depend on $\alpha$, with classical or nonclassical behaviors. This is apparent in Figures 1 and 3 that differ only in the value of $|\alpha|$ and show nonclassical and classical behaviors, respectively. The classical/nonclassical transition value is $|\alpha|_{c} = 0.4539661917$, where $\lim_{\tau \rightarrow\infty} g^{(2)}(\tau) =g^{(2)}(0)$ for $\bar{n}=0.1$ and $r=0.1$.

For the case of the squeezed coherent state $\hat{D}(\alpha) \hat{S}(\xi)|0\rangle$, the condition of nonclassicality (28) becomes
\begin{equation}
\exp \Big{[}\frac{1}{2} |\eta|^2 \Big{(}1- e^{-2r}\Big{)}\Big{]}>1
\end{equation}
for $r>0$,  $\tau=0$, and $\bar{n}=0$. Note that the Gaussian state is less nonclassical than the coherent state $|\xi\rangle$, which corresponds to $\bar{n} =0$. The amount of squeezing is reduced by the factor $(2\bar{n}+1)$, and the squeezing is lost if
$(2\bar{n}+1)>e^{2r}$.

Note that initially at $t=0$, the system is in the thermal state $\hat{\rho}_{0}$, which means that $\alpha=\xi =0$ according to Equations (4) and (5) and so there is no squeezing. However, the squeezing operation is realized by the Hamiltonian (1) thus generating a unitary evolution operation identical to the effect of the squeezing operator $\hat{S}(\xi)$. The squeezing continues indefinitely and so no matter the value of $\bar{n}$, eventually as $\tau$ increases the dynamics will always lead to nonclassical states.

\section{Numerical comparisons}

\begin{figure}
\begin{center}
   \includegraphics[scale=0.3]{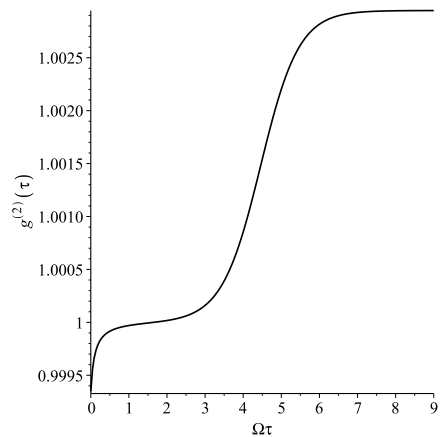}
\end{center}
\label{fig:theFig}
  \caption{Temporal second-order correlation function $g^{(2)}(\tau)$ for $\bar{n}=0.1$, $r=0.1$, and $|\alpha|=5$. Both nonclassical inequalities in (22) are satisfied. One has $g^{(2)}(0) =0.9993$ and $\lim_{\tau\rightarrow \infty } g^{(2)}(\tau)= 1.0029$. The statistics is sub-Poissonian and (26) gives $Q_{M}(0)< 0$.}
\end{figure}

Owing to the equivalence of the nonclassical conditions given by the first of Equation (22) and the Mandel condition $Q_{M}(0)<0$ on the one hand and the equivalence of the nonclassicality condition of the quadrature (25) and condition (29) on the characteristic function, we need study only  numerically the relation of the nonclassical inequalities (22) and (23) for the coherence function $g^{(2)}(\tau)$ and compare them to the nonclassicality criteria (29) for the characteristic function $\chi(\eta)$.

\begin{figure}
\begin{center}
   \includegraphics[scale=0.3]{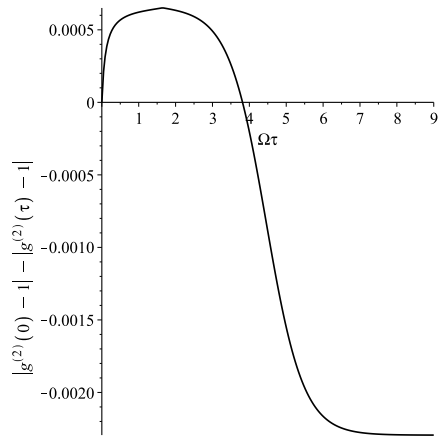}
\end{center}
\label{fig:theFig}
  \caption{Plot of $(|g^{(2)}(0)-1|-|g^{(2)}(\tau)-1|)$ for $\bar{n}=0.1$, $r=0.1$, and $|\alpha|=5$ relevant to inequality (23), which is positive for $0\leq \Omega \tau \leq 3.8151$ and negative for $\Omega \tau > 3.8151$. The former indicates a classical behavior whereas the latter shows nonclassicality.}
\end{figure}

It is interesting that Equation (29) is independent of the coherent parameter $\alpha$ while the coherence function $g^{(2)}(\tau)$ is rather sensitive to the value of $\alpha$. This is so since the dependence of $\chi (\eta)$ on $\alpha$, as given by Equation (9), appears only in the factor $\exp{\big{(}\eta A^*(\tau)- \eta^* A(\tau)\big{)}}$ whose absolute value is unity owing to the argument of the exponential function being a purely imaginary number.

\begin{figure}
\begin{center}
   \includegraphics[scale=0.3]{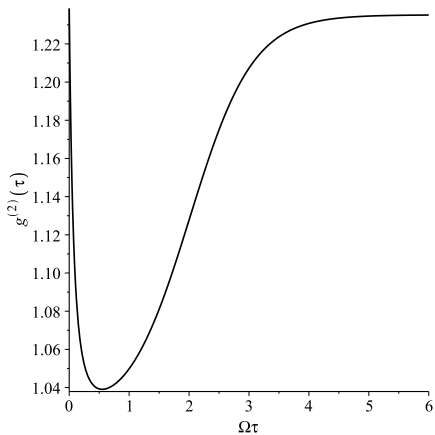}
\end{center}
\label{fig:theFig}
  \caption{Temporal second-order correlation function $g^{(2)}(\tau)$ for $\bar{n}=0.1$, $r=0.1$, and $|\alpha|=0.45$. The behavior of $g^{(2)}(\tau)$ is strictly classical with $g^{(2)}(0) =1.2385$ and $\lim_{\tau\rightarrow \infty } g^{(2)}(\tau)= 1.2352$. Notice the drastic difference from that of Figure 1.}
\end{figure}

Figure 1 shows the strictly nonclassical features of the correlation function $g^{(2)}(\tau)$ since it satisfies the nonclassical inequalities given by Equation (22). On the other hand, Figure 2 shows the mixed classical and quantum features as prescribed by Equation (23) since for $0 \leq \Omega \tau \leq 3.8151$, one has violation of nonclassicality according to (23) whereas for $\Omega \tau >3.8151$ the nonclassical condition (23) is satisfied. The nonclassicality condition (29) is satisfied regardless the value of the coherent parameter $\alpha$, for $\bar{n}=0.1$, $r=0.1$, and $\tau\geq 0$ since $(2\bar{n} +1) e^{-2(\Omega \tau+r)}\leq (2\bar{n} +1) e^{-2 r} = 0.9825<1$. Therefore, the nonclassicality criteria (29) does not prevent the correlation function $g^{(2)}(\tau)$ to show classical behavior.

\begin{figure}
\begin{center}
   \includegraphics[scale=0.3]{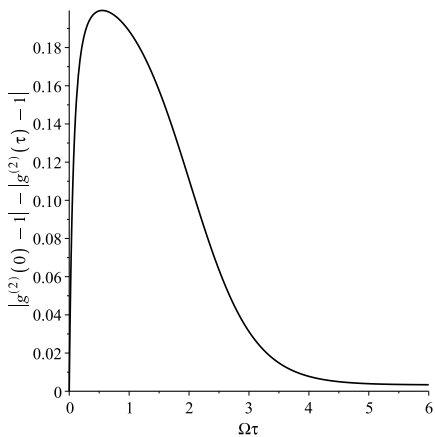}
\end{center}
\label{fig:theFig}
  \caption{Plot of $(|g^{(2)}(0)-1|-|g^{(2)}(\tau)-1|)$ for $\bar{n}=0.1$, $r=0.1$, and $|\alpha|=0.45$ relevant to inequality (23), which is positive for $\Omega \tau \geq 0$.}
\end{figure}

In order to show the strong dependence of the coherence function $g^{(2)}(\tau)$ on the coherent parameter $\alpha$, we show in Figure 3 the strictly classical behavior of $g^{(2)}(\tau)$ for the same values $\bar{n}=0.1$ and $r=0.1$ as those in Figures 1 but with the value of $|\alpha| =0.45$ rather than $|\alpha| =5$ as in Figure 1. In Figure 4 we plot the variable associated with inequality (23) that, together with Figure 3, shows that the system satisfies all the classical inequalities contrary to the nonclassical inequalities (22) and (23). Accordingly, the system behaves classically according to the known inequalities associated with the classical correlation functions $g^{(2)}_{c}(\tau)$; however, homodyne detection measurement of the quadrature-operator expectation values $\Delta x^2_{\lambda}$  would surely show quantum behavior. Thus the statement that (29) is the necessary and sufficient condition for nonclassicality is not reflected in the classical behavior of the correlation function $g^{(2)}(\tau)$.

\begin{figure}
\begin{center}
   \includegraphics[scale=0.3]{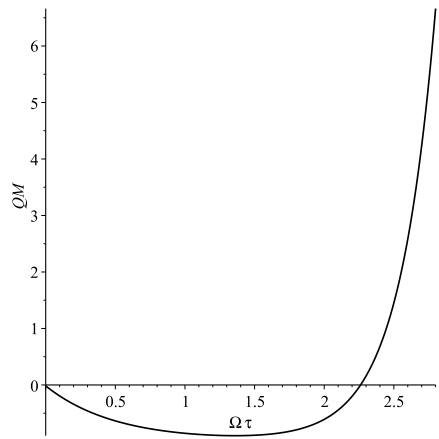}
\end{center}
\label{fig:theFig}
  \caption{Plot of $Q_{M}(\tau)$ for $\bar{n}=0.1$, $r=0.1$, and $|\alpha|=5$. One has $Q_{M}(0)=-0.01636$ and $Q_{M}(\tau)=0$ for $\Omega \tau =2.261$.}
\end{figure}

\begin{figure}
\begin{center}
   \includegraphics[scale=0.3]{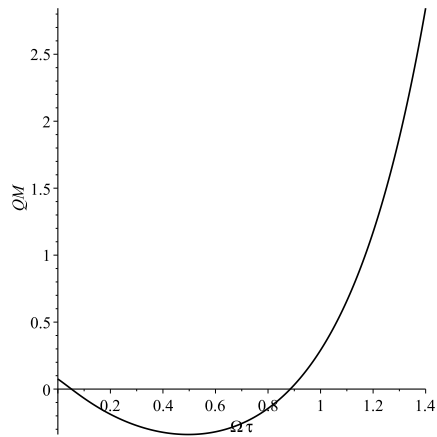}
\end{center}
\label{fig:theFig}
  \caption{Plot of $Q_{M}(\tau)$ for $\bar{n}=0.1$, $r=0.1$, and $|\alpha|=0.45$. One has $Q_{M}(0)=0.07502$ and $Q_{M}(\tau)=0$ for $\Omega \tau = 0.0517, 0.8847$.}
\end{figure}

Finally, Figures 5 and 6 show the temporal behavior of the Mandel parameter $Q_{M}(\tau)$ for the parameters used in Figures 1, 2 and Figures 3, 4, respectively. Recall that the photon number distribution for a coherent field is Poissonian and hence any distribution which is narrower than Poissonian must by necessity correspond to a nonclassical field \cite{GSA13}.

In Figure 5, $Q_{M}(\tau) <0$ for $0< \Omega \tau < 2.261$ and so the field is nonclassical for that range in agreement with results in Figures 1, 2. However, for $\Omega \tau >2.261$, the Mandel criterion indicates a classical behavior contrary to the nonclassical behavior indicated by the coherence function $g^{(2)}(\tau)$ given in Figure 1, which is in agrement with both nonclassical inequalities given in Equation (22).

In Figure 6, nonclassicality is indicated by the Mandel criterion only in the interval $0.0517 < \Omega \tau < 0.8847$, whereas the nonclassicality condition given by (29) is valid for $\tau\geq 0$ for $\bar{n}=0.1$ and $r=0.1$, independent of the value of the coherent parameter $\alpha$.

Again one sees the strong dependence on $|\alpha|$ of the correlation function $g^{(2)}(\tau)$ a dependence that shows both classical and nonclassical behaviors, which is absent in the purported necessary and sufficient nonclassicality criteria (29) for the characteristic function $\chi(\eta)$ owing to (29) being independent of the value of $|\alpha|$.

\section{Summary and discussions}
We calculate the temporal characteristic function (9) and the corresponding field-quadrature (16) and photon-number variances (21) for Gaussian states (2), viz., displaced-squeezed thermal states, where the dynamics is governed solely by the general, degenerate parametric amplification Hamiltonian (1). Our result (9) for the characteristic function is exact and is based on dynamically generating the Gaussian state (2) first from an initial thermal state (3) and subsequently determining the time evolution of the system without assuming statically stationary fields.

We numerically analyze the conditions for nonclassicality as given by the Mandel parameter  $-1\leq G_{M}<0$, squeezing parameter $(\Delta x_{\lambda}^2-1/2)<0$, and characteristic function  $|\chi(\eta)|>1$ and show that the latter condition by itself suffices to determine nonclassicality. The characteristic function nonclassicality condition (29) is contrasted to the violations of the known classical inequalities for the coherence function $g^{(2)}(\tau)$, which violations are given by Equations (22) and (23). Our numerical studies show that the nonclassicality criteria (29), indicating the dynamical state of the system at one time, does not determine the classical/nonclassical behavior of the correlation function $g^{(2)}(\tau)$, which actually represents the measurement of observables of the dynamical system at two different times. We find examples whereby the nonclassicality condition $|\chi(\eta)|>1$ is satisfied while the coherence function $g^{(2)}(\tau)$ satisfies all the known classical conditions. Accordingly, the  necessary and sufficient condition (29) for nonclassicality can be applied only to one-time properties of the system and does not determine the classical or nonclassical nature of two-time properties of the system, for instance, as determined by the coherence function $g^{(2)}(\tau)$.

\appendix*
\section{Second-order coherence}
The degree of second-order temporal coherence is \cite{MA16}
\begin{equation}
g^{(2)}(\tau)= 1 + \frac{n^2(\tau)+ s^2(\tau) +u(\tau) n(\tau) -v(\tau)s(\tau)}{\langle \hat{a}^{\dag}(0)\hat{a}(0)\rangle \langle \hat{a}^{\dag}(\tau)\hat{a}(\tau) \rangle},
\end{equation}
where
\begin{equation}
n(\tau)= (\bar{n}+ 1/2)\cosh \big{(}  \Omega\tau + 2r\big{)} -(1/2)\cosh (\Omega\tau),
\end{equation}

\begin{equation}
s(\tau)= (\bar{n}+ 1/2)\sinh \big{(}\Omega\tau + 2r\big{)} -(1/2)\sinh (\Omega\tau),
\end{equation}

\begin{equation}
u(\tau) = \alpha A^*(\tau) + \alpha^* A(\tau),
\end{equation}
and
\begin{equation}
v(\tau)=\alpha A(\tau)\exp{(-i\theta)}+\alpha^* A^*(\tau)\exp{(i\theta)},
\end{equation}
where $A(\tau)$ is defined by Equation (10).

Equation (10) is the correct expression for $A(\tau)$ and not that given in Ref. 2, where in Equation (13) the purely imaginary number $i$ should not be there. Similarly, there is no $i$ in the square braces of  Equation (A2) in Ref. 2.

\begin{newpage}
\bibliography{}

\begin{thebibliography}{}
\subsection*{References}\label{refs}
\bibitem{GSA13} Agarwal, G.S. \emph{Quantum Optics}, Cambridge University Press: Cambridge, \textbf{2013}.
\bibitem{MA16} Alexanian, M. ``Temporal second-order coherence function for displaced-squeezed thermal states". J. Mod. Opt. \textbf{2016}, 63,  961-967.
\bibitem{WVO99} Welsch, D.-G.; Vogel, W.; Opatrny, T. ``Homodyne detection and quantum state reconstruction", in Progress in Optics, edited by Emil Wolf (North Holland, Amsterdam, 1999), Vol. XXXIX, Cap. II, 63-210.
\bibitem{RL00} Loudon, R. \emph{The Quantum Theory of Light}, Oxford University Press: Oxford, \textbf{2000}.
\bibitem{RC88} Rice, P.R.; Carmichael, H.J. \emph{Quantum IEEE J. Electron.} \textbf{1988}, 24, 1351-1366; Foster, G.T.; Mielke, S.L.; Orozco, L.A. \emph{Phys. Rev. A} \textbf{2000}, 61, 053821.
\bibitem{RV02} Richter. Th.; Vogel, W. Phys. Rev. Lett. \textbf{2002}, 89, 283601.
\end{thebibliography}

\end{newpage}
\end{document}